 \documentclass[reprint,superscriptaddress,amsmath,amssymb,aps,notitlepage,floatfix,longbibliography,dvipdfmx]{revtex4-1}
\usepackage{graphicx}
\usepackage{color}

\usepackage{dcolumn}
\usepackage{upgreek}
\usepackage{gensymb}
\usepackage{float}
\usepackage{bm}
\usepackage{xcolor}

\usepackage{xr}
\makeatletter
\newcommand*{\addFileDependency}[1]{
  \typeout{(#1)}
  \@addtofilelist{#1}
  \IfFileExists{#1}{}{\typeout{No file #1.}}
}
\makeatother

\newcommand*{\myexternaldocument}[1]{
    \externaldocument{#1}
    \addFileDependency{#1.tex}
    \addFileDependency{#1.aux}
}

\myexternaldocument{supps_092820}

\begin{document}
	
\title{Interface enhanced helicity dependent photocurrent in metal/semimetal bilayers}

\author{Hana Hirose}
\affiliation{Department of Physics, The University of Tokyo, Tokyo 113-0033, Japan}

\author{Masashi Kawaguchi}
\affiliation{Department of Physics, The University of Tokyo, Tokyo 113-0033, Japan}

\author{Yong-Chang Lau}
\affiliation{Department of Physics, The University of Tokyo, Tokyo 113-0033, Japan}

\author{Frank Freimuth}
\affiliation{Peter Gr\"{u}nberg Institut and Institute for Advanced Simulation, Forschungszentrum J\"{u}lich and JARA, 52425 J\"{u}lich, Germany}

\author{Masamitsu Hayashi}
\affiliation{Department of Physics, The University of Tokyo, Tokyo 113-0033, Japan}

\newif\iffigure
\figurefalse
\figuretrue

\date{\today}

\begin{abstract}
One of the hallmarks of light-spin interaction in solids is the appearance of photocurrent that depends on the light helicity.
Recent studies have shown that helicity dependent photocurrent (HDP) emerges due to light induced spin current and the inverse spin Hall effect of semimetal thin films.
We have studied HDP in metal/semimetal bilayers.
Compared to Bi single layer films, we find the HDP is enhanced in metal/Bi bilayers.
For the bilayers, the sign of HDP under back illumination reverses from that of front illumination.
The back illumination photocurrent is the largest for Ag/Bi bilayers among the bilayers studied.
Using a diffusive spin transport model, we show that the HDP sign reversal under back illumination is caused by spin absorption and spin to charge conversion at the interface.
Such interfacial effects contribute to the HDP enhancement under front illumination for the bilayers when the Bi layer thickness is small.
These results show that the HDP can be used to assess interface states with strong spin orbit coupling.
\end{abstract}

\maketitle

The exchange of spin angular momentum between electrons plays a fundamental role in modern spintronics as it allows current induced control of magnetism\cite{manchon2019rmp}.
The concept can be extended to interaction of electron spins with light: the transfer of spin angular momentum from light to electrons allow manipulation of magnetization using ultrashort laser pulses in magnetic thin films\cite{kimel2005nature,stanciu2007prl,lambert2014science}.
The interaction of light with electron spin also manifests itself in photocurrents, i.e. currents that flow when light is irradiated to solids.
For example, irradiation of circularly polarized light to materials that possess spin-momentum locked bands results in generation of helicity dependent anisotropic photocurrent. 
The effect, often referred to as the circularly photogalvanic effect (CPGE), has been identified in semiconductor heterostructures\cite{ganichev2002nature,ganichev2004prl}, topological surface states\cite{mciver2012nnano,okada2016prb,pan2017ncomm}, van der Waals structures\cite{yuan2014nnano,ma2017nphys} and (semi-)metallic interface states\cite{hirose2018apl,puebla2019prl}. 
In addition, other forms of photocurrent emerge depending on certain symmetry of the system (e.g. broken structural inversion symmetry and/or broken time reversal symmetry)\cite{zutic2002prl,endres2013ncomm,morimoto2016sciadv,freimuth2017prb,roca2017jjap}. 

\begin{figure}[b]
\begin{center}
\includegraphics[scale=0.55]{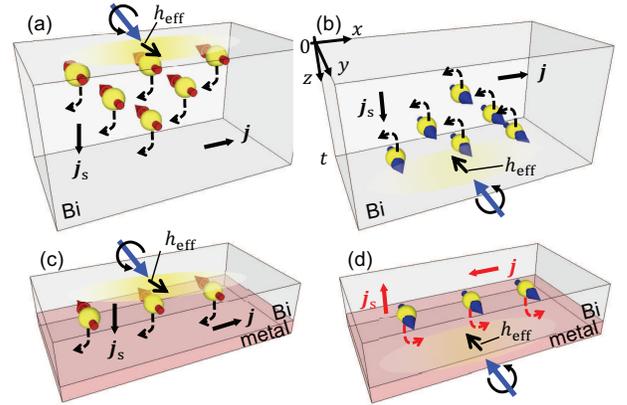}
 \caption{(a-d) Schematic illustrations depicting light induced spin density via the IFE when right-handed circularly polarized light is irradiated to the film from front (a,c) and back (b,d). See Fig.~\ref{fig:setup}(a) for the geometry of front and back illuminations. $h_\mathrm{eff}$ represents the $y$-component of the effective magnetic field associated with circularly polarized light. Gradient in the spin density, caused by the finite light penetration depth, generates a spin current ($\bm{j}_\mathrm{s}$) along the film normal. Spin current is converted to charge current ($\bm{j}$) via the ISHE. The direction to which $\bm{j}_\mathrm{s}$ flows is the same for front and back illuminations for Bi single layer (a,b).  When a spin absorbing interface is present in metal/Bi bilayer (c,d), spin current flows toward the interface. If the film thickness is small (i.e. when the spin density gradient is small), $\bm{j}_\mathrm{s}$ due to interface is dominant and the resulting $\bm{j}$ due to ISHE is opposite to that without the interface under back illumination (compare (b) and (d)). 
}
\label{fig:model}
\end{center}
\end{figure}

Recent studies have shown that helicity dependent photocurrent appears in thin films composed of semimetals (e.g. Bi, doped Bi alloys)\cite{kawaguchi2020condmat}.
The effect has been described assuming that circularly polarized light induces spin density, i.e. an imbalance in the population of carriers with spin parallel and antiparallel to the light spin angular momentum, via the inverse Faraday effect (IFE)\cite{pershan1966pr,hertel2006jmmm,taguchi2011prb,berritta2016prl,qaiumzadeh2016prb,freimuth2016prb,tokman2020prb}.
Due to the finite penetration depth of the light intensity, a gradient in the spin density develops, which causes flow of spin current along the film normal.
The spin current is converted to charge current via the inverse spin Hall effect (ISHE)\cite{saitoh2006apl} of Bi.
This process is sketched in Fig.~\ref{fig:model}(a), which we refer to as the bulk contribution to the helicity dependent photocurrent.

Here we study helicity dependent photocurrent (HDP) in metal/Bi bilayers.
The HDP is found to be larger for Ag/Bi bilayer compared to Bi single layer films.
Front and back light illuminations are used to separate contributions from the bulk and those associated with interface states, if any.
Illustration of the process is described in Fig.~\ref{fig:model}.
For the bulk contribution\cite{kawaguchi2020condmat}, the sign of HDP will be the same for front and back illuminations since the direction of spin current, $\bm{j}_s$ in Figs.~\ref{fig:model}(a) and \ref{fig:model}(b), is the same.
In contrast, the spin direction of the electrons present at the bottom surface of Bi (or the metal/Bi interface for metal/Bi bilayers (Fig.~\ref{fig:model}(c,d))) is opposite for the front and back illuminations. 
Under such circumstance, interfacial effects (e.g. spin absorption, the inverse Rashba-Edelstein effect (IREE)\cite{rojassanchez2014ncomm}) can cause current that flows in opposite direction for the two geometries: compare Figs.~\ref{fig:model}(c) and \ref{fig:model}(d).
Thus the photocurrent measurements using front and back illuminations allow one to separate bulk and interfacial contributions.
We find that the sign of photocurrent reverses for front and back illuminations in Cu/Bi and Ag/Bi bilayers.
The magnitude of the photocurrent under back illumination is the largest for Ag/Bi bilayers, indicating that the degree of spin absorption and spin to charge conversion at the interface is the largest.

Metal/Bi bilayers are deposited on silicon or quartz crystal substrates using RF magnetron sputtering. 
The film structure is sub./seed/$t$ Bi/2 MgO/1 Ta (thickness in nm).
We refer to films with and without the seed layer as seed/Bi bilayer and single Bi layer, respectively. 
The seed layer for the bilayers is 0.5 Ta/2 Cu, 0.5 Ta/2 Ag, 2 W and 0.5 Ta/2 Pt (thickness in nm). 
The 0.5 nm thick Ta layer is deposited before the seed layer to promote its smooth growth.
The 2 MgO/1 Ta layers serve as a capping layer. 
Wires are formed by inserting a metal shadow mask between the substrate and the sputtering target during the deposition process\cite{hirose2018apl}.

\begin{figure}[b]
 \begin{center}
  \includegraphics[scale=0.55]{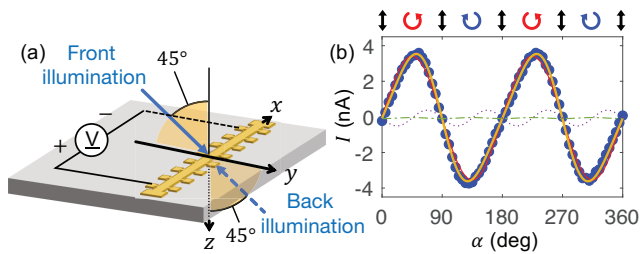}
  \caption{(a) Schematic illustration of the experimental setup. The yellow caterpillar like structure represents the wire made of the film. Definition of front and back illuminations are sketched. 
 (b) The $\alpha$ dependence of the photocurrent ($I$) for Ag/Bi bilayer film with $t\sim65$ nm. The orange solid line shows fit to the data with Eq.~(\ref{eq:fitting}). The red solid, purple dotted and green dashed lines show contributions from the $C$, $L_1$, and $L_2$ terms, respectively.}
  \label{fig:setup}
 \end{center}
\end{figure}

The experimental setup and definition of the coordinate axis are described in Fig.~\ref{fig:setup}(a). 
Light is irradiated from an oblique angle ($\sim45 ^\circ$) to the wire.
We refer to front and back illuminations when light is irradiated to the wire from the film side or from the back of the substrate, respectively.
The light plane of incidence is always orthogonal to the wire's long axis.
A continuous wave semiconductor laser with wavelength $\lambda$ and power $P$ is used as the light source.
Typical results from $\lambda=405$ nm and $P\sim2.5$ mW are presented.
The laser spot size is $\sim$ 0.5 mm in diameter.
The photovoltage of the wire (width: $w\sim0.4$ mm, length: $L\sim7$ mm) is measured while illuminating light through a quarter wave plate.
The angle ($\alpha$) of the quarter wave plate's optical axis with respect to the light plane of incidence defines the light helicity: the light is linearly polarized when $\alpha=0^{\circ},90^{\circ},180^{\circ}$, $270^{\circ}$ and circularly polarized when $\alpha=45^{\circ},225^{\circ}$ (left handed) and $135^{\circ},315^{\circ}$ (right handed). 
The photovoltage is converted to photocurrent by dividing the voltage with the resistance of the wire inside the laser spot ($\sim$0.5 mm long).

Figure~\ref{fig:setup}(b) shows the $\alpha$ dependence of the photocurrent from a Ag/Bi bilayer under front illumination. 
The data is fitted to the following function to extract parameters with different symmetries:
\begin{equation}
\begin{aligned}
    I =& C\sin{2(\alpha + \alpha_0)}\\
     &+L_1\sin{4(\alpha + \alpha_0)}+L_2\cos{4(\alpha + \alpha_0)}+I_\mathrm{0}
\label{eq:fitting}
\end{aligned}
\end{equation}
$C$ represents photocurrent that depends on the helicity of light, whereas $L_1$ an $L_2$ reflects photocurrent that differs for circularly and linearly polarized light\cite{mciver2012nnano,okada2016prb}.
$I_0$ corresponds to an offset photocurrent that does not depend on $\alpha$ and $\alpha_0$ is an offset angle associated with the experimental setup (here $\alpha_0 \sim 1^\circ$).
The fitted curve is shown by the orange solid line in Fig.~\ref{fig:setup}(b), which agrees well with the data.
As evident, the photocurrent is dominated by the helicity dependent term ($C$): the other contributions ($L_1$, $L_2$ and $I_0$) are typically smaller than $C$; see Appendix section~\ref{sec:L1} and Fig.~\ref{fig:L1L2I0}.

\begin{figure}[b]
\centering
\includegraphics[scale=0.6]{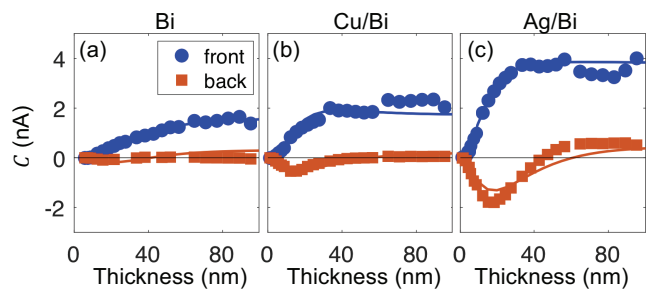}
\caption{(a-c) Bi layer thickness ($t$) dependence of the HDP ($C$) under front (blue circles) and back (orange squares) illuminations for Bi single layer (a), Cu/Bi bilayer (b), Ag/Bi bilayers (c). The solid lines show calculated $C$ using the model.}
\label{fig:bilayer}
\end{figure}

The Bi layer thickness dependence of $C$ for Bi single layer, Cu/Bi and Ag/Bi bilayers using front and back light illuminations are shown in Figs.~\ref{fig:bilayer}(a-c).
For front illumination, all structures show an increase in $C$ with increasing Bi layer thickness ($t$) until it saturates. 
The saturation value of $C$ is the largest for the Ag/Bi bilayer and is the smallest for Bi single layer.
For back illumination, $C$ is nearly zero for Bi single layer in the entire thickness range studied.
Note that in this system (Bi single layer) $C$ is also close to zero under front illumination when $t \lesssim 10$ nm.
We infer that the Bi layer within this thickness do not contribute to the generation of light induced spin density, thus forming a spin excitation dead layer possibly due to difference in the structure/texture of Bi.
If the thickness of the dead layer, defined as $t_\mathrm{d}$, is close to the light penetration depth, light irradiated from the back of the substrate will not reach a region where a non-zero spin density can be induced, resulting in near zero $C$ for back illumination.
In contrast, for Cu/Bi and Ag/Bi bilayers, a large negative $C$ is found (under back illumination) when $t$ is small.
$C$ increases with increasing $t$ and changes its sign from negative to positive when $t \sim 40$ nm. 
These results show that there are two competing effects that contribute to the generation of HDP in the bilayers.

To model the system, we solve the spin diffusion equation with a source term associated with light induced spin density. 
We first define the source term. 
The number of photons absorbed in Bi at position $z$ is defined as $n_\mathrm{ph}(z)$: $z=0$ correspond to the top surface of the Bi layer in contact with the MgO/Ta capping layer; see Fig.~\ref{fig:model}(b).
We assume the following simplified functional form for $n_\mathrm{ph}(z)$: 
\begin{equation}
\begin{aligned}
	n_\mathrm{ph}(z) = A \exp{ (- \alpha_\mathrm{eff} z)}.
\label{eq:nph}
\end{aligned}
\end{equation}
(For back illumination, substitute $(t-z)$ for $z$.)
$\alpha_\mathrm{eff}$ is the effective extinction constant and $A$ is a coefficient representing the light amplitude for a given Bi layer thickness $t$. 
In the absence of multiple reflections within the film, $A$ is a constant and equals the power of the incident light.
Here we take into account multiple reflection and thus $A$ depends on $t$.

To obtain $A$ and $\alpha_\mathrm{eff}$, we measure the reflectivity $R$ and transmittance $T$ of circularly polarized light irradiated to films.
The measured $R$ and $T$ for both front and back illuminations for Bi single layer, Cu/Bi and Ag/Bi bilayers are shown in Appendix Fig.~\ref{fig:abscoeff}.
Taking into account multiple reflections that take place at the substrate/film and film/air interfaces (see Appendix section~\ref{sec:multiple} and Fig.~\ref{fig:multiple_ref}), we fit the thickness dependence of $R$ and $T$ to estimate the refractive index $n$ and extinction coefficient $\kappa$ of the films. 
The estimated values of $n$ and $\kappa$ are shown in Table~\ref{table:fitpar}, which are in good agreement with past report on bulk Bi\cite{werner2009jpcrd}.

The absorbance ($P_\mathrm{a}$) of a film is expressed as
\begin{equation}
\begin{aligned}
P_\mathrm{a} = P (1 - R - T),
 \label{eq:Pa}
\end{aligned}
\end{equation}
where $P$ is the light power incident on the film.
We equate $P_\mathrm{a}$ divided by the energy of one photon $\frac{hc}{\lambda}$ ($c$ is the speed of light, $h$ is the Planck constant) and the area of a laser spot $S$ with the thickness integrated sum of $n_\mathrm{ph}(z)$, i.e. 
\begin{equation}
\begin{aligned}
    P_\mathrm{a} \left(\frac{hc}{\lambda}\right)^{-1} \frac{1}{S} &= \int_0^t n_\mathrm{ph}(z) dz \\
    &=
    \int_0^t A \exp{ (- \alpha_\mathrm{eff} z)} dz.
\label{eq:Pabs}
\end{aligned}
\end{equation}
Using Eqs.~(\ref{eq:Pa}) and (\ref{eq:Pabs}) and $R$ and $T$ calculated using the experimentally obtained $n$ and $\kappa$ (see Appendix Eqs.~(\ref{eq:RTfront}) and (\ref{eq:RTback}) for the relation between $R$, $T$ and $n$, $\kappa$), we estimate $A$ for each $t$.
We assume $\alpha_\mathrm{eff} = \frac{4 \pi \kappa}{\lambda}$ for all samples.
Substituting $A$ and $\alpha_\mathrm{eff}$ into Eq.~(\ref{eq:nph}), we obtain the number of photons ($n_\mathrm{ph}(z)$) absorbed at $z$ for a given film thickness $t$.
For simplicity, we assume one photon absorbed at position $z$ in the film generates spin density equivalent of $\frac{\hbar}{2}$ ($\hbar \equiv h / (2 \pi)$).
Defining $n_\mathrm{s}$ as the number of electrons that are spin polarized along the light spin angular momentum, i.e. the spin density, we obtain
\begin{equation}
\begin{aligned}
	n_\mathrm{s}(z)= n_\mathrm{ph}(z).
\label{eq:lightspinconv}
\end{aligned}
\end{equation}
We use $n_\mathrm{s}(z)$ as the source term of the spin diffusion equation.

The spin diffusion equation is expressed using the chemical potential difference ($\mu_\mathrm{s}$) of the electrons with spin pointing parallel and antiparallel to the light spin angular momentum, that is,
\begin{equation}
\begin{aligned}
    \frac{\partial^2 \mu_\mathrm{s}(z)}{\partial z^2}=\frac{1}{\lambda_\mathrm{s}^2}\mu_\mathrm{s}(z)+ \frac{e^2}{\sigma_{xx}} n_\mathrm{s}(z),
\label{eq:spindiffusionmodel}
\end{aligned}
\end{equation}
where $\lambda_\mathrm{s}$ is the spin diffusion length and $\sigma_{xx}$ is the conductivity of Bi ($\sigma_{xx} \sim 1 \times 10^5 \ (\Omega \cdot \mathrm{m})^{-1}$).
We solve Eq.~(\ref{eq:spindiffusionmodel}) to obtain $\mu_\mathrm{s}(z)$, which can be converted to spin current density ($\bm{j}_\mathrm{s}$) via the relation,
\begin{equation}
\begin{aligned}
    \bm{j}_\mathrm{s}(z) = - \frac{\sigma_{xx}}{2e} \nabla \mu_\mathrm{s}(z)
\label{eq:js}
\end{aligned}
\end{equation}
$j_{\mathrm{s},i}$ represents spin current along the $i$-direction with polarization pointing along the light spin angular momentum, which we represent by a unit vector $\bm{e}_\sigma$.
The boundary condition is defined as
\begin{equation}
\begin{gathered}
    j_{\mathrm{s},z}(z=0) = 0,\ \ 
    j_{\mathrm{s},z}(z=t) = 0,
\label{eq:bc}
\end{gathered}
\end{equation}
for Bi single layer and 
\begin{equation}
\begin{gathered}
	j_{\mathrm{s},z}(z=0) = 0,\ \ 
    j_{\mathrm{s},z}(z=t) = -\frac{\sigma_{xx}}{2e} \frac{\mu(z=t)}{l_\mathrm{int}},
\label{eq:bc_back}
\end{gathered}
\end{equation}
for the seed/Bi bilayers.
The seed/Bi interface at $z=t$ is assumed to absorb spin current.
The degree of absorbance is characterized by $l_\mathrm{int}$\cite{rojassanchez2014ncomm}.
Note that this boundary condition does not explicitly include contributions from, for example, the IREE: it simply describes the presence of an interface that varies the spin current boundary condition.
Finally, the ISHE converts the spin current to charge current density ($\bm{j}_\mathrm{c}$):
\begin{equation}
\begin{aligned}
    \bm{j}_\mathrm{c}(z)=\theta_\mathrm{SH} \bm{j}_\mathrm{s}(z) \times \bm{e}_\sigma,
\label{eq:jc}
\end{aligned}
\end{equation}
where $\theta_\mathrm{SH}$ is the spin Hall angle.
The total charge current along $x$ ($I_{\mathrm{c,}x}$), which is measured experimentally, is obtained by integrating the $x$ component of $\bm{j}_{c}$ over $z$ and multiplying $\sin(\pi / 4)$ (to account for the oblique incidence of light) and the width of the wire. 


\begin{figure}[t]
\begin{center}
\includegraphics[scale=0.55]{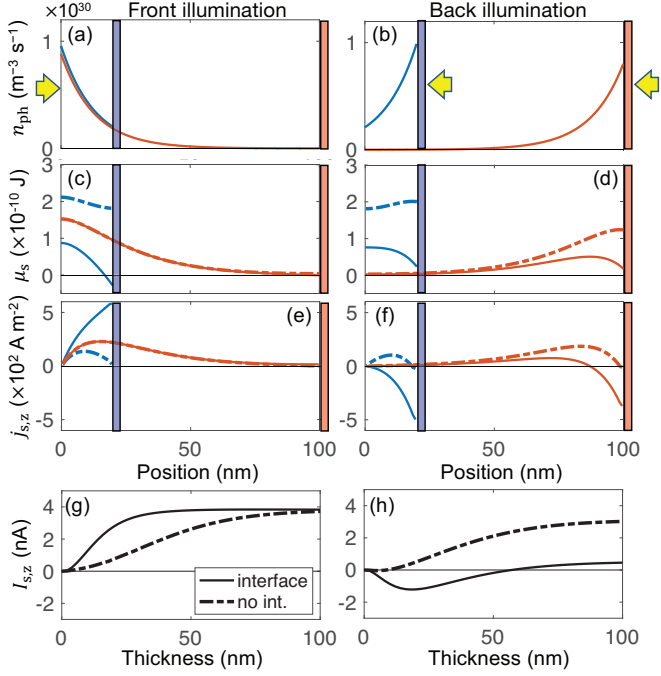}
 \caption{(a-f) Number of absorbed photons ($n_\mathrm{ph}$) (a,b), chemical potential difference ($\mu_\mathrm{s}$) of the electrons with spin pointing parallel and antiparallel to the light spin angular momentum  (c,d), and the spin current density ($j_{\mathrm{s},z}$) along $z$ (e,f), plotted against film position at $z$ for front (a,c,e) and back (b,d,f) illuminations.
 The blue and red lines show results when the film thickness ($t$) is 20 nm and 100 nm, respectively. The red and blue thick vertical lines indicate the position of metal/Bi interface and the yellow arrow represents the direction from which light is irradiated. (g,h) $t$ dependence of the $z$-component of the thickness integrated spin current $\big(I_{\mathrm{s},z} = w \int_0^t  j_{\mathrm{s},z}(z) dz \big)$ for front (g) and back (h) illuminations. (a-h) The solid and dashed lines represent calculation results when a spin absorbing interface is present and absent, respectively. The parameters used are the same with those of Ag/Bi bilayers described in Table~\ref{table:fitpar}.}
\label{fig:cal}
\end{center}
\end{figure}

Figures~\ref{fig:cal}(a-f) shows the calculated $n_\mathrm{ph}(z)$, $\mu_\mathrm{s}(z)$ and $j_{\mathrm{s},z}(z)$ for films with $t=20$ nm (blue lines) and $t=100$ nm (red lines) under front and back illuminations.
The solid and dashed lines show results with ($\frac{1}{l_\mathrm{int}} \neq 0$) and without ($\frac{1}{l_\mathrm{int}} = 0$) a spin absorbing interface, respectively.
The difference of the maximum $n_\mathrm{ph}$ among all conditions shown in Figs.~\ref{fig:cal}(a,b) is due to multiple reflections within Bi.
The profiles of $\mu_\mathrm{s}$ and $j_{\mathrm{s},z}$ significantly change when the interface is present. 

The spin current integrated across the Bi layer thickness, $I_{\mathrm{s},z} = w \int_0^t  j_{\mathrm{s},z}(z) dz$, is plotted as a function of $t$ in Figs.~\ref{fig:cal}(g,h).
Without the interface (dashed lines), $I_{\mathrm{s},z}$ increases with increasing $t$ until saturation and is positive for front and back illuminations. 
With the interface (solid lines), however, the signs of $I_{\mathrm{s},z}$ for front and back illuminations are opposite when $t$ is small.
$I_{\mathrm{s},z}$ for back illumination is negative for small $t$ and changes its sign a $t \sim 50$ nm.
As evident in Fig.~\ref{fig:cal}(f) (see also Fig.~\ref{fig:model}(c) for a schematic illustration), the presence of spin absorbing interface induces spin current toward the interface.
For films with small $t$, $I_{\mathrm{s},z}$ is dominated by the back flow toward the interface over contribution from the gradient in spin density (i.e. gradient of $\mu_\mathrm{s}$).
In addition, $I_{\mathrm{s},z}$ for front illumination with the interface is larger compared to that without the interface.

With this model, we fit the experimental results of front and back light illuminations simultaneously, with $\theta_\mathrm{SH}$ and $\lambda_\mathrm{s}$ used as the fitting parameters.
For Cu/Bi and Ag/Bi bilayers, $l_\mathrm{int}$ is also varied to fit the data (for Bi single layer, $1/l_\mathrm{int} = 0$).
A phenomenological spin excitation dead layer ($t_d$) is introduced, which is determined by the thickness which causes near zero $C$ for small $t$ under front illumination.
The calculated $C$ are shown by the blue and orange solid lines in Figs.~\ref{fig:bilayer}(a-c), which show good agreement with the experimental results.
The parameters used for the calculations are listed in Table~\ref{table:fitpar}.

$\theta_\mathrm{SH}$ of Bi obtained here is considerably larger than that of previous reports estimated using spin pumping measurements\cite{hou2012apl,emoto2016prb}, but is in relatively good agreement with that of sputtered Bi-rich BiSb alloys\cite{chi2020sciadv}. 
The estimated $\theta_\mathrm{SH}$ of Ag/Bi bilayers is nearly twice as large as that of other structures. 
The resistivity of Bi in the bilayers studied is similar, suggesting that the spin Hall angle of Bi takes similar value\cite{chi2020sciadv}.
In the model, we assume the interface modifies the spin current profile in Bi via changes in the boundary condition. The IREE\cite{rojassanchez2014ncomm,nomura2015apl,karube2016apex} can convert the spin current that flows into the interface to generate a charge current, providing an additional channel.
Interestingly, the size of $l_\mathrm{int}$ required to describe the results for Ag/Bi bilayers is close to that reported in similar systems\cite{rojassanchez2014ncomm}.
With the current model, however, it is difficult to separate contributions from the ISHE and IREE due to the back flow of spin current to the interface (note that the back flow occurs due to the presence of spin absorbing interface).



\begin{table}[t]
 \caption{Parameters used in the model calculations.}
 \label{table:fitpar}
 \centering
  \begin{tabular}{c@{\hspace{0.3cm}}r@{\hspace{0.3cm}}c@{\hspace{0.3cm}}c@{\hspace{0.3cm}}c@{\hspace{0.3cm}}c@{\hspace{0.3cm}}c}
   \hline \hline
   & $\theta_\mathrm{SH}$ & $\lambda_\mathrm{s}$ & $n$ & $\kappa$ & $l_\mathrm{int}$ & $t_d$\\\
   Structure & & (nm) & & & (nm) & (nm) \\
   \hline 
   Bi single layer & 0.7 & 18 & 1.3 & 3.2 & N/A & 12 \\
   Cu/Bi bilayer &  0.7  & 18 & 1.3 & 3.0 & 6 & 5\\
   Ag/Bi bilayer & 1.4 & 18 & 1.3 & 2.5 & 3 & 1\\
   \hline
  \end{tabular}
\end{table}

We have also studied photocurrent in W/Bi and Pt/Bi bilayers, in which the seed layer exhibits significantly larger spin Hall effect than Cu and Ag\cite{hoffmann2013ieee,sinova2015rmp}; see Appendix Fig.~\ref{fig:WPt}. 
We find that $C$ is not particularly large for these bilayers compared to that of Cu/Bi and Ag/Bi bilayers. 
As the carrier density of Bi is more than three orders of magnitude smaller than that of typical metals (W and Pt)\cite{kawaguchi2020condmat}, we consider spin current flowing into the seed layer hardly contributes to the charge current within the seed layer via the ISHE\cite{fert2001prb}.

In summary, we have studied bulk and interface contributions to the helicity dependent photocurrent (HDP) in metal/Bi bilayers. 
As reported previously, the bulk contribution originates from the ISHE of Bi, which converts light induced spin current to charge current within Bi.
In metal/Bi bilayers, we find that not only the HDP increases under front illumination, compared to Bi single layer, but also the sign of HDP reverses when light is illuminated from the back.
Using a diffusive spin transport model, we show that the metal/Bi interface acts as a strong spin sink and modifies the profile of spin current in Bi. 
Such change in the spin current profile results in an enhancement of HDP due to the ISHE of Bi as well as the IREE at the interface. 
We find the largest HDP in Ag/Bi bilayers, both under front and back illuminations, suggesting strong contributions from the interface. 
These results thus demonstrate means to study spin absorption and spin to charge conversion at interfaces using circularly polarized light.
Given that the photocurrent in metal/Bi bilayers is dominated by the helicity dependent component, the large HDP found here can be exploited for polarization sensitive detectors in optical communications\cite{liu2020nmat} as well as light spin angular momentum detectors for quantum optics\cite{togan2010nature}.

\begin{acknowledgments}
This work was partly supported by JST CREST (JPMJCR19T3), JSPS Grant-in-Aids (JP15H05702, JP16H03853), Yamada Science Foundation and the Center of Spintronics Research Network of Japan. Y.-C.L. is supported by JSPS International Fellowship for Research in Japan (JP17F17064). H.H. acknowledge financial support from Materials Education program for the future leaders in Research, Industry, and Technology (MERIT).
\end{acknowledgments}

\section{Appendix}

\subsection{\label{sec:L1} Helicity and polarization dependent photocurrent}
\begin{figure}[h]
 \begin{center}
  \includegraphics[scale=0.6]{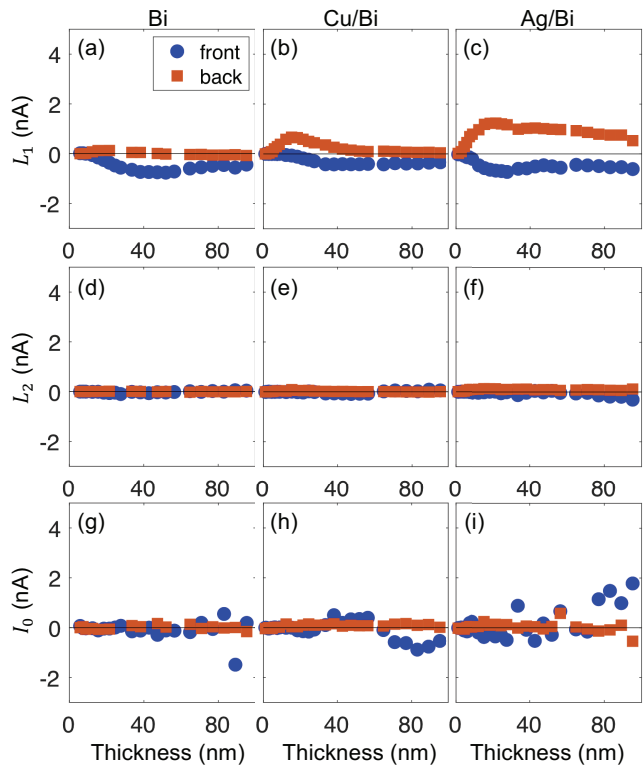}
  \caption{(a-l) Bi layer thickness $t$ dependence of the photocurrent components, $L_1$ (a,b,c), $L_2$ (d,e,f) and $I_0$ (g,h,i), obtained by fitting the data with Eq.~(\ref{eq:fitting}). The results show photocurrent from Bi single layer (a,d,g), Cu/Bi bilayer (b,e,h) and Ag/Bi bilayer (c,f,i). Blue circles and orange squares represent results under front and back illuminations, respectively. }
  \label{fig:L1L2I0}
 \end{center}
\end{figure}

Components of the photocurrent, $L_1$, $L_2$ and $I_0$, obtained by fitting the data ($I$ vs. $\alpha$) with Eq.~(\ref{eq:fitting}), are shown in Fig.~\ref{fig:L1L2I0} as a function of Bi layer thickness ($t$) for Bi single layer, Cu/Bi and Ag/Bi bilayers.
In all cases, $C$ (see Fig.~\ref{fig:bilayer}) and $L_1$ show a similar thickness dependence, suggesting that the two effects have a common origin.
$L_2$ and $I_0$ are negligible in all structures.
For the thicker Bi films, $I_0$ shows relatively large fluctuation, which we consider is related to laser induced heating effects that may originate from the large thermo-electric effects of Bi. 


\subsection{\label{sec:multiple} Multiple reflection model and absorption of light}
We calculate the reflectivity $R$ and transmittance $T$ of the system assuming that multiple reflections takes place at the top and bottom interfaces of the semimetal (Bi) layer.
We model the system using three media: air (medium 1), the film including the seed and capping layers (medium 2), and the quartz substrate (medium 3).
Since the seed layer and the capping layer are thin compared to the light wavelength, we include them as part of the Bi layer.
Note that the Cu and Ag seed layers reduce the amplitude of light transmission.
We have measured the light transmission probability ($T_\mathrm{seed}$) of a 0.5 Ta/2 Cu deposited on quartz crystal and found $T_\mathrm{seed} \sim 0.76$.
We assume 0.5 Ta/2 Ag possesses similar $T_\mathrm{seed}$.

\begin{figure}[h!]
 \begin{center}
  \includegraphics[scale=0.65]{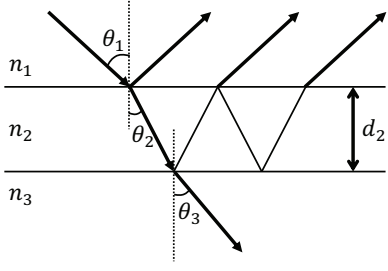}
  \caption{Schematic illustration of the multiple reflection that takes place within the film.}
  \label{fig:multiple_ref}
 \end{center}
\end{figure}

The refractive index of the three media is defined as $n_1=1.0 + i 0$ (air), $n_2 = n + i \kappa$ (film) and $n_3= 1.5 + i 0$ (substrate). 
The thickness of the film is $d_2$.
The other interface of the substrate, substrate/air, is treated as a transmission loss. 
Transmission loss of the substrate is studied separately using a substrate without the film.
The transmission probability is $T_\mathrm{sub}^s \sim 0.908$ for $s$-polarized light and $T_\mathrm{sub}^p \sim 0.991$ for $p$-polarized light.
Light is irradiated from an oblique angle of $45^{\circ}$.
Schematic illustration of the system is shown in Fig.~\ref{fig:multiple_ref}.





When a $s(p)$-polarized light with wavelength $\lambda$ is incident from medium 1 (air) with an oblique angle $\theta_1$ on medium 2 (film), the amplitude of the reflected light ($r_{123}^{s(p)}$) and the transmitted light ($t_{123}^{s(p)}$) are
\begin{equation}
\begin{aligned}
    r_{123}^{s(p)} &= \frac{r_{23}^{s(p)}+r_{12}^{s(p)} \exp{(i \gamma)}}{1+ r_{23}^{s(p)} r_{12}^{s(p)} \exp{(i \gamma)}},\\
    t_{123}^{s(p)} &= \frac{t_{23}^{s(p)} t_{12}^{s(p)} \exp{(i \gamma /2)}}{1+ r_{23}^{s(p)} r_{12}^{s(p)} \exp{(i \gamma)}}.
\label{eq:rt20}
\end{aligned}
\end{equation}
$r_{ij}^{s(p)}$ and $t_{ij}^{s(p)}$ are the Fresnel reflection and transmission coefficients for $s(p)$-polarized light, defined as
\begin{equation}
\begin{aligned}
    r_{ij}^s &= \frac{n_i \cos\theta_i - n_j \cos \theta_j}{n_i \cos\theta_i + n_j \cos \theta_j},\\
    t_{ij}^s &= \frac{2 n_i \cos\theta_i}{n_i \cos\theta_i + n_j \cos \theta_j},\\
    r_{ij}^p &= \frac{n_j \cos\theta_i - n_i \cos \theta_j}{n_j \cos\theta_i + n_i \cos \theta_j},\\
    t_{ij}^p &= \frac{2 n_i \cos\theta_i}{n_j \cos\theta_i + n_i \cos \theta_j}.
\label{eq:rt_ij}
\end{aligned}
\end{equation}
In Eq.~(\ref{eq:rt20}), $\gamma \equiv \frac{4 \pi}{\lambda} n_2 d_2 \cos{\theta_2}$, where $\theta_2$ is the complex refraction angle in the film.
The reflection ($R_{123}^{s(p)}$) and transmission ($T_{123}^{s(p)}$) probabilities of the $s(p)$-polarized light are written as
\begin{equation}
\begin{aligned}
    R_{123}^{s(p)} &= \left| r_{123}^{s(p)} \right|^2,\\
    T_{123}^{s(p)} &= \left| \frac{n_3 \cos{\theta_3}}{n_1 \cos{\theta_1}} \right| \left| t_{123}^{s(p)} \right|^2,
\label{eq:RT20}
\end{aligned}
\end{equation}
where $\theta_3$ is the complex refraction angle of the transmitted light in medium 3.


The reflection $R_{123}^c$ and transmission $T_{123}^c$ probabilities of a circularly polarized light are expressed as
\begin{equation}
\begin{aligned}
    R_{123}^c &= \frac{1}{2}(R_{123}^s+R_{123}^p),\\
    T_{123}^c &= \frac{1}{2} (T_{123}^s T_\mathrm{sub}^s +T_{123}^p T_\mathrm{sub}^p ) T_\mathrm{seed}.    
\label{eq:RTfront}
\end{aligned}
\end{equation}
The absorbance of the film for circularly polarized light is calculated as
\begin{equation}
\begin{aligned}
    A_{123}^c = \frac{1}{2} \big( \{1-(R_{123}^s+T_{123}^s)\} + \{1-(R_{123}^p+T_{123}^p)\} \big).
\label{eq:Pabsorptionfront}
\end{aligned}
\end{equation}

For back illumination, we exchange parameters of medium 1 with those of medium 3.
One needs to replace Eqs.~(\ref{eq:RTfront}) and (\ref{eq:Pabsorptionfront}) with the following relations:
\begin{equation}
\begin{aligned}
    R_{123}^c &= \frac{1}{2}\big( (T_\mathrm{sub}^s)^2 R_{123}^s+(T_\mathrm{sub}^p)^2 R_{123}^p \big) \ \ \mathrm{[back \ illumination],}\\
    T_{123}^c &= \frac{1}{2} (T_{123}^s T_\mathrm{sub}^s +T_{123}^p T_\mathrm{sub}^p ) T_\mathrm{seed} \ \ \mathrm{[back \ illumination],}
\label{eq:RTback}
\end{aligned}
\end{equation}
\begin{equation}
\begin{aligned}
A_{123}^c &= \frac{1}{2} \big( T_\mathrm{seed} T_\mathrm{sub}^s \{1-(R_{123}^s+T_{123}^s)\}\\
&+ T_\mathrm{seed} T_\mathrm{sub}^p \{1-(R_{123}^p+T_{123}^p)\} \big)  \ \ \mathrm{[back \ illumination].}
\label{eq:Pabsorptionback}
\end{aligned}
\end{equation}
\\


\subsection{\label{sec:opticalconstants} Measurements of the optical constants}

\begin{figure}[h!]
 \begin{center}
  \includegraphics[scale=0.6]{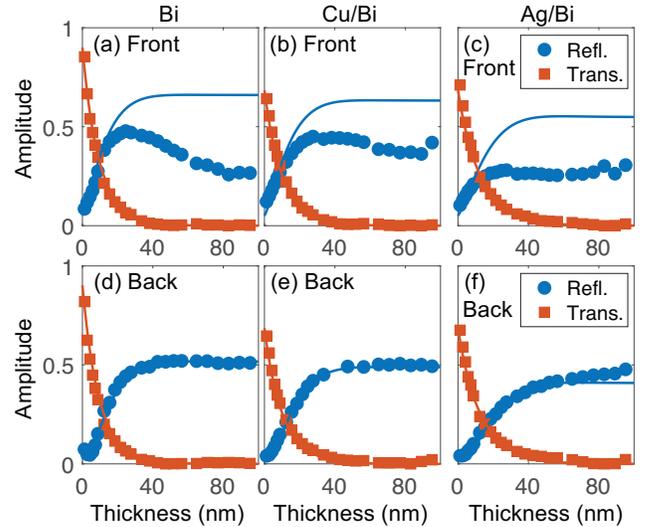}
  \caption{
  (a-c) Reflectivity $R$ (filled markers) and transmittance $T$ (open markers) of left handed circularly polarized light under front illumination (upper panels) and back illumination (lower panels) for Bi single layer (a), Cu/Bi bilayer (b) and Ag/Bi bilayer (c), plotted as a function of $t$. Solid lines show the calculated values of $R$ and $T$, respectively, that best fit the experimental results.
  }
  \label{fig:abscoeff}
 \end{center}
\end{figure}

The optical constants of the films are estimated from measurements of the reflectivity ($R$) and transmittance ($T$) of circularly polarized light. 
The measured $R$ and $T$ for front and back illuminations for Bi single layer, Cu/Bi and Ag/Bi bilayers are shown by the symbols in Figs.~\ref{fig:abscoeff}(a-c) and \ref{fig:abscoeff}(d-f), respectively.
The $t$ dependence of $R$ and $T$ are fitted with Eqs.~(\ref{eq:RTfront}) and (\ref{eq:RTback}) to extract $n$ and $\kappa$ of the film.
The extracted values are listed in Table~\ref{table:fitpar}.
The absorbance are calculated using Eqs.~(\ref{eq:Pabsorptionfront}) and (\ref{eq:Pabsorptionback}).
$A_{123}^c$ is equivalent to $P_\mathrm{a}$ in Eq.~(\ref{eq:Pabs}).

\subsection{\label{sec:WPt} Helicity dependent photocurrent for W/Bi and Pt/Bi bilayers}

\begin{figure}[h!]
 \begin{center}
  \includegraphics[scale=0.6]{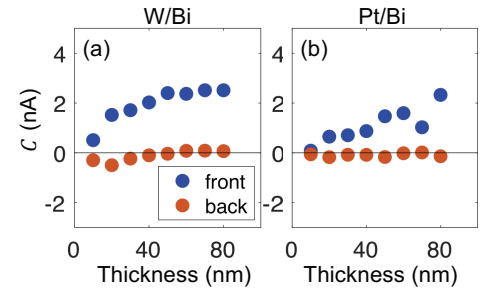}
  \caption{
  (a,b) $t$ dependence of the HDP ($C$) under front (blue circles) and back (orange squares) illuminations for W/Bi (a) and Pt/Bi bilayers (b).}
  \label{fig:WPt}
 \end{center}
\end{figure}

The $t$ dependence of $C$ for W/Bi and Pt/Bi bilayers are shown in Fig.~\ref{fig:WPt}.
 
\bibliography{ref_100920}

\end{document}